\documentclass[aps,prl,twocolumn,nobibnotes]{revtex4-2}  
\usepackage{dcolumn}
\usepackage{graphicx}
\usepackage{color}
\usepackage{amssymb}
\usepackage{amsmath}
\usepackage{bm}
\usepackage[colorlinks=true,linktocpage=true,breaklinks=true]{hyperref}
\usepackage{multirow}
\usepackage{braket}
\usepackage{appendix}

\begin{document}
\title{Interaction-Induced Topological Phase Transition in Magnetic Weyl Semimetals}
\author{Konstantinos Sourounis}
\email{konstantinos.sourounis@univ-amu.fr}
\author{Aurélien Manchon}
\email{aurelien.manchon@univ-amu.fr}
\affiliation{Aix-Marseille Université, CNRS, CINaM, Marseille, France}
\begin{abstract}
Despite the tremendous interest raised by the recent realization of magnetic Weyl semimetals and the observation of giant anomalous Hall signals, most of the theories used to interpret experimental data overlook the influence of magnetic fluctuations, which are ubiquitous in such materials and can massively impact topological and transport properties. In this work, we predict that in such magnetic topological systems, the interaction between electrons and magnons substantially destabilizes the Weyl nodes, leading to a topological phase transition below the Curie temperature. Remarkably, the sensitivity of the Weyl nodes to electron-magnon interaction depends on their spin chirality. We find that Weyl nodes with a trivial chirality are more sensitive to electron-magnon interactions than Weyl nodes presenting an inverted chirality, demonstrating the resilience of the latter compared to the former. Our results open perspectives for the interpretation of the transport signatures of Weyl semimetals, especially close to the Curie temperature.
\end{abstract}

\maketitle

\textit{Introduction} - Magnetic topological materials, both insulators \cite{hasan2010colloquium,qi2011topological,tokura2019magnetic} and semimetals \cite{armitage2018weyl,bernevig2022progress}, are currently under scrutiny as they can emulate high-energy physics such as the quantized magnetoelectric effect associated with axions \cite{qi2008topological,essin2009magnetoelectric,Liu2020} and the chiral anomaly associated with Weyl fermions \cite{zyuzin2012topological,son2013chiral}. While the original realization of topological insulators was made by doping nonmagnetic topological insulators with magnetic impurities \cite{chang2013experimental,chang2015high}, the identification and synthesis of intrinsic magnetic topological materials have made massive progress in the last few years \cite{morali2019,Belopolski2019,Liu2019d,otrokov2019prediction,deng2020quantum,Xu2020c,belopolski2021signatures}. With the realization of magnetic Weyl semimetals \cite{bernevig2022progress}, there has been extensive theoretical work on their classification based on symmetries \cite{Xu2020,Elcoro2021,robredo2024new} and their anomalous transport properties \cite{manna2018colossal,Noky2020,Samanta2023}. One of the most intriguing hallmarks of these materials is the giant anomalous Hall effect associated with the Weyl nodes, detected in ferromagnetic topological semimetal candidates \cite{Liu2018,kim2018large} with cubic Heusler alloys taking the spotlight \cite{li2020giant,singh2021anisotropic,Guin2021,Singh2024extended,Chatterjee2023nodal}. Whereas the topological transport properties of such systems have been studied thoroughly at the zero-temperature limit, the influence of magnetic fluctuations, ubiquitous at finite temperatures, has been largely overlooked.

Determining the topology of materials in the presence of interactions is a challenging topic as interactions break the well-defined nature of electronic quasiparticles \cite{wan2011topological,witczak2012topological,kuroda2017evidence,yin2018giant,belopolski2021signatures,soldini2023}. While the correct definition of topology in an interacting system is an open problem \cite{Zhao2023failure,Gavensky2023connecting,setty2023electronic}, the most widely used topological diagnosis is the $N_3$-number \cite{wang2012simplified,gurarie2011single,volovik2003universe}, which converges towards the Chern number in the noninteracting limit. To date, most attention has been paid to materials whose topological properties are governed by electron correlations, such as topological Mott \cite{witczak2014interacting,morimoto2016weyl,roy2017,crippa2020nonlocal,wagner2023mott}, Kondo \cite{lai2018weyl,chen2024emergent} or fractional insulators \cite{bollmann2023topological} and semimetals. One particularly interesting proposal is that the electron-phonon interaction can induce a topological phase transition upon changing the temperature, as first proposed in Ref. \cite{Garate2013,Saha2014Phonon} and later confirmed by {\em ab initio} studies \cite{Monserrat2016Temperature,Antonius2016Temperature,monserrat2019unraveling,chen2024temperature}. In this situation, the electron mass is enhanced by electron-phonon coupling, resulting in a temperature-dependent orbital gap, leading to the topological phase transition. In magnetic topological materials, where the topology is intimately related to the spin chirality of the wave function, the electron-magnon interaction is expected to induce interband transitions that can deteriorate the Berry curvature and, thereby, impact the topology. This interaction is crucial to properly understand the giant anomalous Hall signature recently reported in magnetic Weyl semimetals such as $\mathrm{Co_3Sn_2S_2}$ \cite{morali2019,Belopolski2019,Liu2019d,Xu2020c} or (Ni,Pt)MnSb \cite{Singh2024extended}.

\begin{figure}[b]
\includegraphics[width=8.6cm,height=8.6cm,keepaspectratio]{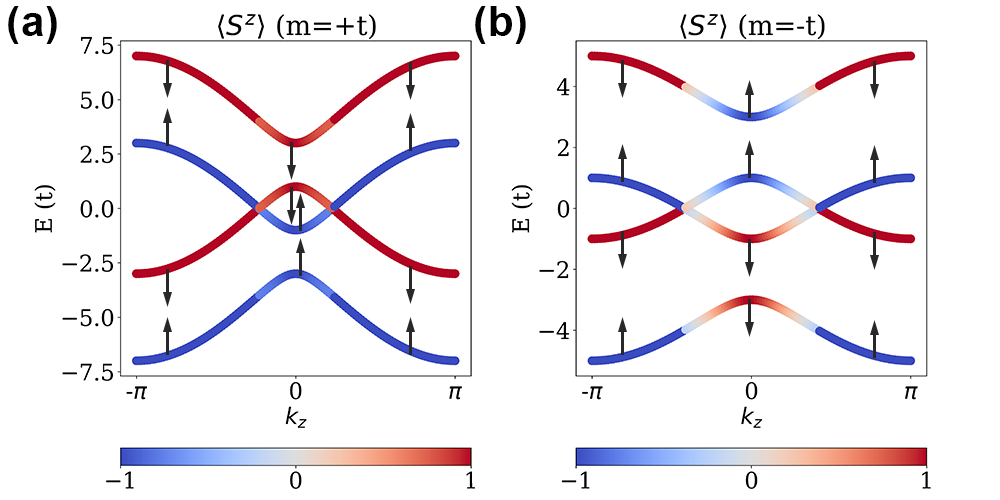}
\caption{The band structure of the magnetic Weyl semimetal projected on the spin chirality at $T=0$K. The arrows indicate the direction of the energy shift induced by the electron-magnon self-energy, as explained in the main text. Two different regimes are considered: (a) the trivial ($m=+t$) and (b) the inverted ($m=-t$) one.}
\label{fig:fig1}
\end{figure}

In this Letter, we show how the electron-magnon interaction can control the topological phase of a magnetic topological semimetal and heavily impact its transport properties. Remarkably, we find that in a magnetic Weyl semimetal, this interaction enables temperature-dependent topological phase transitions below the Curie temperature, and we demonstrate that the sensitivity of the Weyl nodes to the electron-magnon interaction depends on the chirality of the nodes.

\textit{Electron-Magnon Interaction} - The magnetic Weyl semimetal is modeled by a four-band Hamiltonian with both spin ($\sigma=\uparrow,\downarrow$) and orbital ($\eta =A, B$) degrees of freedom, representing a magnetic cubic Heusler material \cite{Garate2013, Rauch2017model}. This model was used specifically to explain experimental features of Co$_2$Mn(Ga,Al) \cite{Noky2020} and Fe$_2$(Co,Ni)(Ga,Al) compounds \cite{Mende2021}. The total Hamiltonian is ${\cal H}={\cal H}^e+{\cal H}^m+{\cal H}^{em}$, where ${\cal H}^e$ and  ${\cal H}^m$ represent the electronic and magnetic systems, and ${\cal H}^{em}$ their interaction. The electronic Hamiltonian reads
\begin{eqnarray}
{\cal H}_{\bf k}^e&=&\boldsymbol{d_k\cdot\sigma}\otimes\tau^x+M_k\sigma^0\otimes\tau^z+\Delta\langle S^z \rangle \sigma^z\otimes\tau^0,\label{eq:1}
\end{eqnarray}
where $\tau^i$ and $\sigma^i$ are Pauli matrices in the orbital and spin space respectively, $\tau^0,\sigma^0$ are the identity matrices, $d_k^i=-2\lambda \sin{k_i}$ represent the spin-orbit coupling along the $i$-directions ($i=x, y, z$), and $M_k=m+2t(3-\sum_{i}\cos{k_i})$ is the mass parameter. $\Delta$ is the exchange between itinerant and localized electrons and $\langle S^z \rangle$ is the expectation value of the localized electron spins.

In addition to the electron-magnon interaction, magnon-magnon interaction should also be considered, as it leads to quasiparticle renormalization and magnetization quenching. To do so, several flavors of perturbation theory can be implemented, which presents a computationally demanding task (see, e.g., \cite{Tiwari2021}). To assess the influence of electron-magnon interaction over the full temperature range, it is sufficient to consider a phenomenological temperature dependence of the spin number, $ \langle S^z\rangle =S(1-T/T_c^\beta)^\alpha$ \cite{Wang2022}. Here, we choose $\beta=1$, $\alpha =1/3$ and the Curie temperature $T_c=8J$, $J$ being the magnetic exchange. This approach captures qualitatively the physics of the magnons across the range of $T$=0K to $T=T_c$, as recently shown in two- and three-dimensional systems \cite{Tiwari2021,Pavizhakumari2025}.

In the absence of magnetization ($\Delta=0$), this Hamiltonian represents a Dirac semimetal. By tuning the magnetization, the system adopts two distinct phases, an insulating one (I), $\Delta<|m|$, and a Weyl semimetal phase (WSM), $\Delta>|m|$. In the I phase,  the electronic bands are insulating in all directions, while in the WSM phase, two Weyl Points exist in the $k_z$ direction as depicted in Fig. \ref{fig:fig1}. In addition, switching the sign of $m$ changes the spin chirality of Weyl nodes, as illustrated by the color scale in Fig. \ref{fig:fig1}. In the following, we refer to $m>0$ ($m<0$) as the trivial (inverted) regime. When the chemical potential is set to zero, $\mu=0$, this model is equivalent to a stack of 2D Chern Insulators along the $k_z$ direction \cite{Burkov2011}. In this case, the anomalous Hall conductivity is proportional to the distance between the two Weyl points along this axis $\sigma_{xy}=(e^2/\pi h)\Delta_{WP}$, where  $\Delta_{WP}$ is the distance in momentum space of the Weyl points, similar to its two-band version \cite{armitage2018weyl}.

Let us now turn our attention to the magnetic system, described by 
\begin{eqnarray}
{\cal H}^m&=&-J\sum_{\langle ij\rangle} \boldsymbol S_i\cdot \boldsymbol S_j+K\sum_i (S_i^z)^2,\label{eq:2}
\end{eqnarray}
where $J$ ($>$0) is the nearest-neighbor exchange interaction and $K$ is the uniaxial anisotropy, taken perpendicular to the plane. In the presence of magnetic fluctuations, this Hamiltonian can be readily transformed in the magnon basis via Holstein-Primakoff transformation, as described in the Supplemental Materials \cite{SuppMat}. Finally, the interaction between the magnetic fluctuations and the itinerant electrons is given by
\begin{eqnarray}
{\cal H}^{em}&=&\Delta\sqrt{\frac{1}{2}}\sum_i (S_i^+\sigma_-+ S_i^-\sigma_+)\otimes\tau^0.\label{eq:3}
\end{eqnarray}
Applying the Holstein-Primakoff transformation, the electron-magnon coupling can be rewritten in the second quantization representation
\begin{eqnarray}
   {\cal H}^{em}_{\bf k} =\Delta\sqrt{2S}\sum_{\bf q,\eta}\bigg(a_{\bf q}c_{\eta\downarrow ,\bf k+\bf q}^{\dagger}c_{\eta\uparrow ,\bf k}+a^{\dagger}_{-\bf q}c_{\uparrow\eta ,\bf k+\bf q}^{\dagger}c_{\eta\downarrow ,\bf k}\bigg).\notag \\\label{eq:4}
\end{eqnarray}
Here $a_{\bf q}$ is the magnon annihilation operator, and $c_{\eta\sigma ,\bf k}$ is the annihilation operator for an electron with orbital $\eta$ and spin $\sigma$. Since the eigenstates of Eq. \eqref{eq:1} are spin- and orbital-mixed, spin-flip processes induced by the electron-magnon scattering further mix the bands, thereby modifying their topology. 
To assess the impact of electron-magnon interaction on the band structure, we compute the electron self-energy using quantum field theory \cite{mahan2013many} and obtain
\begin{eqnarray}
    \hat{\Sigma}^{R/A}_{\bf k,\eta\downarrow,\eta'\downarrow}(\epsilon,T) &=&  \sum_{\bf q,\nu} \Phi_{\nu,\bf k+\bf q}^{\eta\downarrow,\eta'\downarrow} \frac{n_F(\epsilon_{\nu,\bf k+\bf q})+n_B(\omega_{\bf q})}{\epsilon\pm i0^+-\epsilon_{\nu,\bf k+\bf q}+\omega_{\bf q}},\\
    \hat{\Sigma}^{R/A}_{\bf k,\eta\uparrow,\eta'\uparrow}(\epsilon,T) &=&  \sum_{\bf q,\nu}\Phi_{\nu,\bf k+\bf q}^{\eta\uparrow,\eta'\uparrow} \frac{1-n_F(\epsilon_{\nu,\bf k+\bf q})+n_B(\omega_{\bf q})}{\epsilon\pm i0^+-\epsilon_{\nu,\bf k+\bf q}-\omega_{\bf q}},\notag\\
    \quad
\end{eqnarray}
where $n_F$ $(n_B)$ is the Fermi-Dirac (Bose-Einstein) distribution of the electrons (magnons). The matrix elements $\Phi_{\nu,\bf k}^{\eta\sigma,\eta'\sigma'}$ and the calculation of self-energies are given explicitly in the Supplementary Material \cite{SuppMat} (see also references  \cite{nolting2009quantum,woolsey1970electron,jo2021visualizing,maeland2021electron} therein). For the temperature range that we consider, only the distribution of the magnons is affected, and the distribution of electrons can be approximated by a step function. The numerical broadening $0^+$ is set to 0.1$t$.
\begin{figure}[t]
\includegraphics[width=8.6cm,height=8.6cm,keepaspectratio]{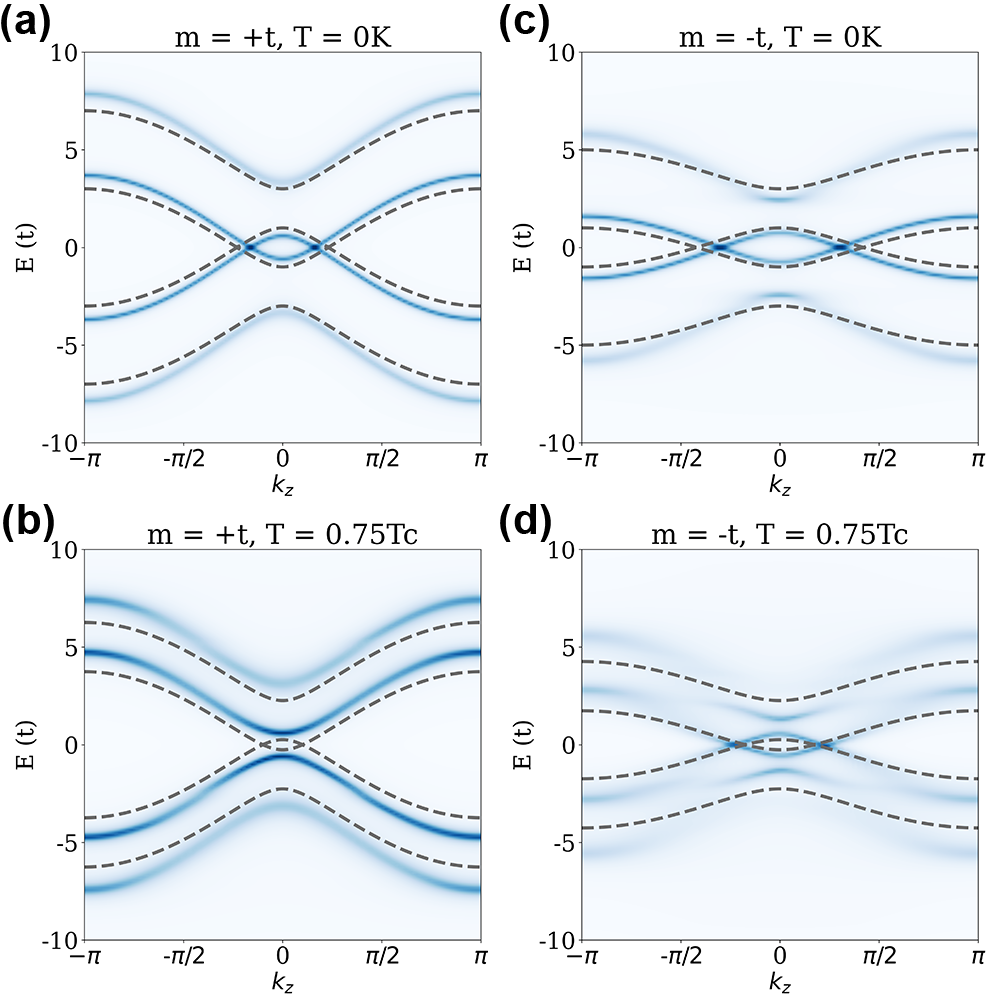}
\caption{The spectral function of the electrons (color) given by Eq. \eqref{eq:7} versus the non-interacting spectrum (dashed lines)  for (a) $m=+t$, $T=0$K, (b) $m=+t$, $T=0.75T_c$, (c) $m=-t$, $T=0$K, (d) $m=-t$, $T=0.75T_c$. The other parameters are set to $t=1$, $\Delta=2t$, $\lambda = t$, $S=1$, $J=t/100$, $K=J/10$.}
\label{fig:fig2}
\end{figure}

The electron-magnon interaction in a ferromagnetic metal causes the emission (absorption) of a magnon and the annihilation (creation) of an electron with spin down (up). This interaction causes a softening of the spin splitting of the ferromagnet, which, in the absence of spin-orbit coupling, can be written for each sublattice $\eta$ as, $\Delta_{eff}^{\eta}=2\Delta+\Sigma_{\eta,\downarrow}-\Sigma_{\eta,\uparrow}$ \cite{hertz1973electron}. In the trivial case, $m=+t$, electrons in the ferromagnetic Weyl model have a similar spin polarization as a ferromagnetic metal, as depicted in Fig. \ref{fig:fig1} (a). As such, it is expected to accelerate the softening of the magnetization, $\langle S^z\rangle(T)$, together with the magnon-magnon interaction mentioned above. In the inverted regime, $m=-t$, the spin chirality is reversed, as depicted in Fig. \ref{fig:fig1} (b), which causes the increase of the effective magnetization $\Delta_{eff}$ in the region between the Weyl points, in competition with the effective softening. Overall, the effect is balanced, and the Weyl points survive at higher temperatures than in the trivial regime.

We first evaluate the ability of the interaction to impact the electrons' spectrum at zero and finite temperatures. To do so, we calculate the one-electron spectral function 
\begin{equation}
    A_{\bf k}(\epsilon)=-\frac{1}{\pi}{\rm Im(Tr}[{G}^R_{\bf k}(\epsilon)]), \label{eq:7}
\end{equation}
where the Green's function is in the Bloch basis, as discussed in \cite{SuppMat}, $G_{\bf k}^{R/A}(\epsilon)=(\epsilon\pm i\Gamma-{\cal H}_{\bf k}^e-\Sigma_{\bf k}^{R/A}(\epsilon,T))^{-1}$, where $\Gamma$ is a homogeneous broadening due, e.g., to impurity scattering. Figure \ref{fig:fig2} displays the interacting spectral function (color) compared to the non-interacting spectrum of the electrons (grey dashed lines). Throughout this work, we adopt the set of parameters: $t=1$, $\Delta=2t$, $\lambda = t$, $S=1$, $J=t/100$, $K=J/10$, and two different cases $m=\pm t$ \cite{SuppMat}. These parameters guarantee that at the Fermi level, only two Weyl points exist in the $k_z$ direction, and there is no other contribution to the anomalous Hall effect other than the topological one, as in Refs. \cite{Noky2020,Rauch2017model,Mende2021}.

For $T=0$K [Fig. \ref{fig:fig2}(a,c)], the only contribution of the electron-magnon interaction is the zero-point fluctuation, and at $T=0.75T_c$ [Fig. \ref{fig:fig2}(b,d)], the thermally excited magnons greatly enhance the interaction. We can see in Fig. \ref{fig:fig2}(a) and (c) that for both $m=\pm t$ cases, the interactions bring the Weyl points closer to each other compared to the non-interacting case, which indicates a weaker topology and a reduced anomalous Hall transport. However, at higher temperatures [Fig. \ref{fig:fig2}(b,d)], once the thermal magnons take over, the two cases diverge. For $m=+t$, the Weyl points are annihilated by the interactions, {\em before} the Curie temperature is reached. This result demonstrates that the topology is highly sensitive to temperature, and the geometrical contribution to the Hall transport is expected to vanish. For $m=-t$, the location of the Weyl points is only slightly affected by the temperature, indicating that the topological features may remain resilient even at high temperatures. Nevertheless, the bands are strongly broadened due to interactions, which can potentially impact transport by increasing diffusion.

\textit{Interacting Topology and Transport} - To assess how these interactions affect the topology and Hall transport of the magnetic Weyl semimetal, we first compute the two-dimensional many-body topological number \cite{wang2012simplified,witczak2014interacting}
\begin{eqnarray}
    &&N_3(k_z)= \frac{e^2}{h}\int d\eta\int dk_xdk_y\times\notag\\ &&{\rm Tr}[G_{\bf k}^T(\partial_{\eta}G_{\bf k}^{-1})G^T_{\bf k}(\partial_{k_x}G_{\bf k}^{-1})G_{\bf k}^T(\partial_{k_y}G_{\bf k}^{-1})- (x\leftrightarrow y)],\notag\\\label{eq:8}
\end{eqnarray}
which is  resolved in the $k_z$ direction and where $\hat{G}^T_{\bf k}(\eta)=(i\eta-{\cal H}_{\bf k}^e-\Sigma_{\bf k}^T(\eta,T))^{-1}$ is the topological Green's function. To convert the retarded Green's functions and self-energies into the topological ones, we make the substitution $\epsilon+i\Gamma\rightarrow i\eta$, where the integration in Eq. \eqref{eq:8} is over $\eta$. The advantage of this expression is that it accounts for changes in the spectrum, geometry, and broadening of the quasiparticles. This number is equivalent to the non-interacting Chern number in two dimensions, and it allows us to evaluate the topology at each $k_z$ point.

\begin{figure}[t]
\includegraphics[width=8.6cm,height=8.6cm,keepaspectratio]{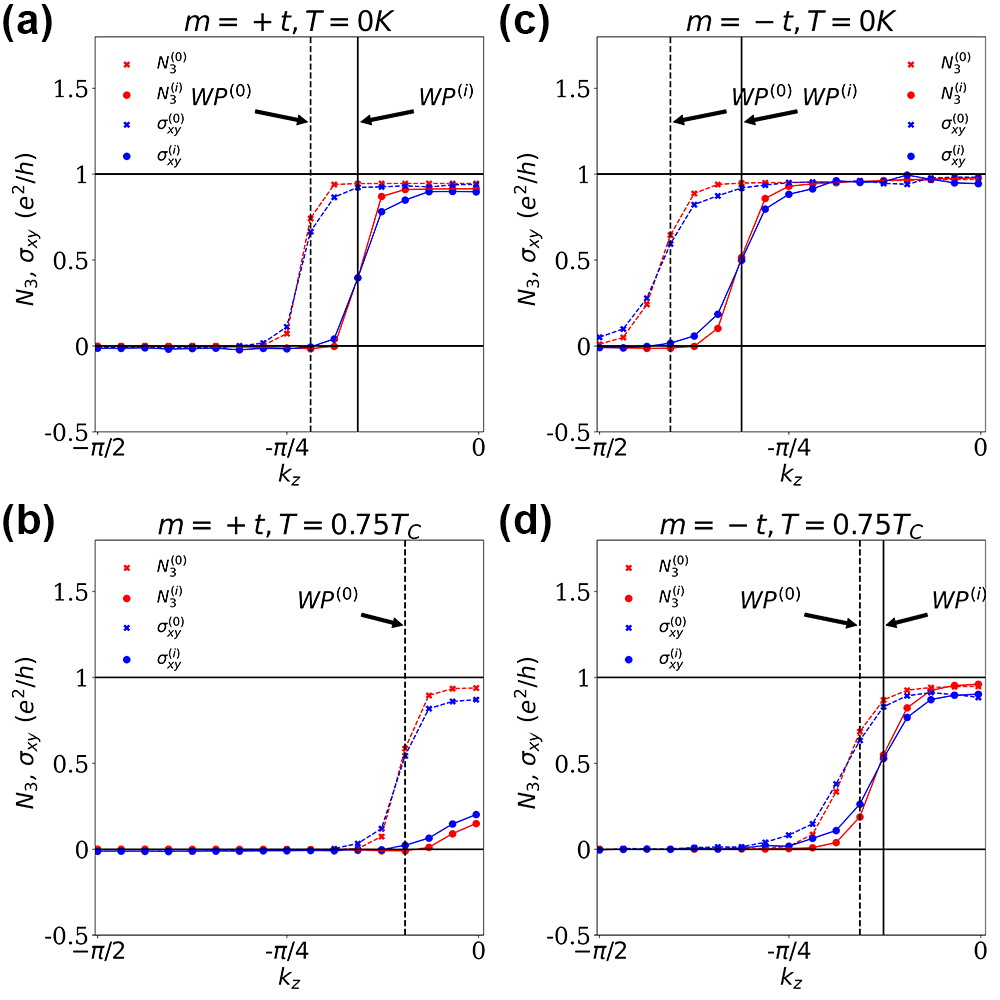}
\caption{The non-interacting and interacting many-body topological number, $N_3^{(0)}$ and $N_3^{(i)}$ [see Eq. \eqref{eq:8}], and the corresponding anomalous Hall conductivity, $\sigma_{xy}^{(0)}$ and $\sigma_{xy}^{(i)}$ [see Eq. \eqref{eq:9}], for the same configuration as in Fig. 2: (a) $m=+t$, $T=0$K (b) $m=+t$, $T=0.75T_c$ (c) $m=-t$, $T=0$K (d) $m=-t$, $T=0.75T_c$. The Weyl points, $WP^{(0,i)}$, are indicated by the vertical lines. We set $\Gamma=0.2t$ in Eq. \eqref{eq:9}.}
\label{fig:fig3}
\end{figure}

In addition, to further confirm the topological nature of the interacting picture, we compute the intrinsic contribution to the anomalous Hall conductivity via Green's function method \cite{bonbien2020symmetrized} (see Supplemental Material \cite{SuppMat} and references  \cite{giustino2017,katsnelson2008,maeland2021electron} therein)
\begin{eqnarray}
    &\sigma_{xy}&(k_z)=\frac{e^2}{h}\int d\epsilon n_F(\epsilon)\int dk_xdk_y\times\notag\\
    &\rm{Tr}&\{\partial_{k_x}G_{\bf k}^{-1}(G_{\bf k}^R-G_{\bf k}^A)\partial_{k_y}G_{\bf k}^{-1}(\partial_{\epsilon}G_{\bf k}^R+\partial_{\epsilon}G_{\bf k}^A)-(x\leftrightarrow y)\}\notag\\\label{eq:9}
\end{eqnarray}
where $\partial_{k_i}G_{\bf k}^{-1}=-(\partial_{k_i}{\cal H}_{\bf k}^e+\partial_{k_i}\Sigma_{\bf k}^R(\eta,T))$ and $\partial_{\epsilon}G_{\bf k}^{R/A}=-(G_{\bf k}^{R/A})\cdot(1-\partial_{\epsilon}\Sigma_{\bf k}^{R/A}(\epsilon,T))\cdot(G_{\bf k}^{R/A})$. For simplicity, we have considered $\partial_{\epsilon/\eta}\Sigma_{\bf k}(\epsilon,T)\approx 0$.  Equation \eqref{eq:9} is the Fermi sea contribution to the transport, equivalent to the anomalous Hall conductivity due to Berry curvature in the non-interacting limit. 

For the magnetic Weyl semimetal model we employ, it is known that the Hall conductivity in the $z$-direction is proportional to the distance between the Weyl points, the length of the Fermi arc \cite{armitage2018weyl}, of the system, which gives a direct connection between conductivity and topology. In Fig. \ref{fig:fig3}, we calculate the many-body topological number $N_3$ using Eq. \eqref{eq:8} (red symbols) and the anomalous Hall conductivity $\sigma_{xy}$ using Eq. \eqref{eq:9} (blue symbols), for the same configurations as in Fig. \ref{fig:fig2}. 

In the absence of interactions (superscripts (0), dashed lines), the values of both $N_3$ and $\sigma_{xy}$ tend to be quantized to 1 for $k_z$ inside the topological region (i.e., at the right-hand side of $WP^{(0)}$) and zero outside it (at the left-hand side of $WP^{(0)}$). In the presence of interactions, the many-body calculations confirm the observations from the spectral properties in Fig. \ref{fig:fig2}. In Fig. \ref{fig:fig3}(a) and (c), in both regimes ($m=\pm t$), the distances of the Weyl points (solid vertical lines) are closer than the non-interacting calculation (dashed vertical lines). When the temperature increases, the results of the two regimes differ. For $m=+t$ [Fig. \ref{fig:fig3}(b)], we see that interactions destroy the topology at high temperatures, and both $N_3$ and $\sigma_{xy}$ vanish in this limit. For $m=-t$ [Fig. \ref{fig:fig3}(d)], both topological and transport properties survive even at high temperatures. 

\begin{figure}[t]
\includegraphics[width=8.6cm,height=8.6cm,keepaspectratio]{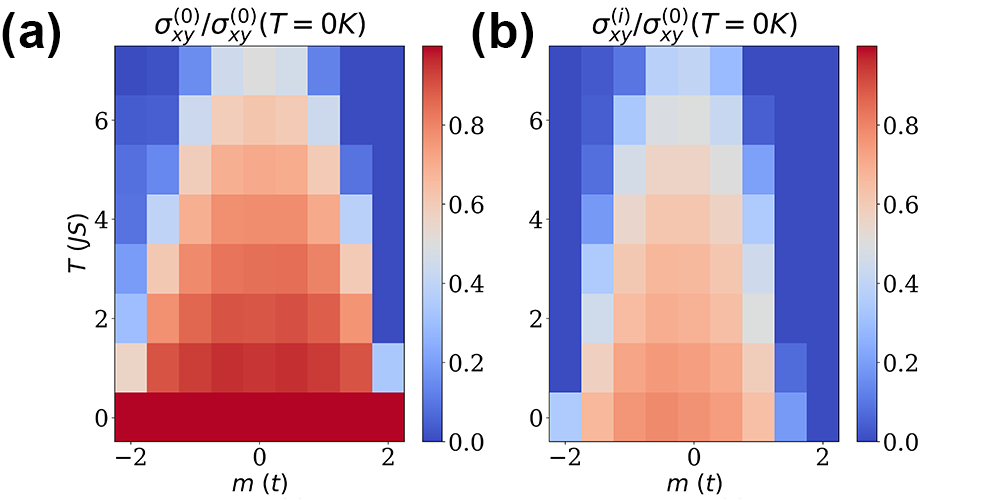}
\caption{The phase diagram of the anomalous Hall conductivity as a function of temperature from Eq. \eqref{eq:9} of the (a) non-interacting (0) and (b) the interacting system (i) for different values of the mass of orbitals $m$. The magnetization quenching due to magnon-magnon interaction is accounted for in both cases $\langle S^z \rangle$.}
\label{fig:fig4}
\end{figure}

In Fig. \ref{fig:fig4}, we plot the phase diagram of the anomalous Hall conductivity [Eq. \eqref{eq:9}] as a function of the temperature and the mass term $m$, with and without interactions. We compute the ratio between the conductivity at temperature $T$ and the conductivity at zero temperature. In Fig. \ref{fig:fig4}(a), we consider only the influence of the magnetization quenching due to magnon-magnon interaction, $\sigma_{xy}^{(0)}(T)/\sigma_{xy}^{(0)}(T=0{\rm K})$, whereas in Fig. \ref{fig:fig4}(b), we include the influence of the electron-magnon interaction, $\sigma_{xy}^{(i)}(T)/\sigma_{xy}^{(0)}(T=0{\rm K})$. In the absence of interactions [Fig. \ref{fig:fig4} (a)], the trivial ($m>0$) and inverted regimes ($m<0$) exhibit a similar temperature dependence. Close to $m=0$, the conductivity vanishes at the Curie temperature as the system goes from a ferromagnetic to a paramagnetic state. Away from $m=0$, the conductivity vanishes at lower temperatures as the system transits toward the insulating phase. In the presence of interactions [Fig. \ref{fig:fig4} (b)], the picture is different. First, at zero temperature, the conductivity is substantially reduced compared to the noninteracting case due to the onset of zero-point magnetic fluctuations. Furthermore, the phase diagram is more asymmetric. In the trivial regime ($m>0$), the electron-magnon interaction cooperates with the magnon softening, causing a decrease in the conductivity at lower temperatures than in the noninteracting case [Fig. \ref{fig:fig4} (a)]. In the inverted regime ($m<0$), the two effects compete, as discussed above, and the conductivity survives at higher temperatures. Thus, the presence of the inverted spin chirality enforces the resilience of the topological state.

\textit{Discussion} - Our predictions can shed light on the recent experiments performed on $\mathrm{Co_2Mn(Ga,Al)}$ \cite{Belopolski2019,li2020giant}, $\mathrm{Co_3Sn_2S_2}$ \cite{morali2019,Liu2019d,Xu2020c,sun2021polarized}, or (Ni,Pt)MnSb \cite{Singh2024extended}. The precise nature of the electron-magnon interaction in these materials requires careful consideration, as certain compounds—$\mathrm{Co_3Sn_2S_2}$ and NiMnSb—are predicted to be half-metals, with only one spin channel at the Fermi level, which would inherently suppress electron-magnon scattering \cite{katsnelson2008}. These calculations are usually conducted at $T$=0K and in the absence of spin-orbit coupling; turning on the spin-orbit coupling leads to the loss of the half-metallicity, corroborating recent experiments \cite{Howlader2020}. Remarkably, temperature-dependent spin-resolved optical spectroscopy has recently revealed that the Weyl points in $\mathrm{Co_3Sn_2S_2}$ are destroyed at $T$=131K, below the Curie temperature ($T$=176K) \cite{sun2021polarized}. In NiMnSb, a half-metallic ferromagnet at the Fermi level \cite{katsnelson2008}, the transport was recently attributed to crossings away from the Fermi level \cite{Singh2024extended}, where both spin chiralities exist. Since the impact of the electron-magnon interaction depends on the spin chirality, one can reasonably expect that different spots in the band structure may behave differently upon increasing the temperature. In $\mathrm{Co_2Mn(Ga,Al)}$, spin polarization near the Weyl points \cite{Sumida2020} and enhanced magnetic damping \cite{charles2019} manifest the existence of substantial electron-magnon interactions.

Another important open question is the coexistence of interactions that can potentially impact the topology and transport as the temperature increases. While both electron-phonon and electron-magnon interactions transfer momentum to a bosonic bath, only the electron-magnon interaction induces spin flips, thereby altering the chirality of the eigenstates. Importantly, the electron-phonon interactions and electron-magnon interactions {\em compete} with each other: electron-phonon interactions favor the Weyl semimetal phase in the trivial regime ($m>0$) \cite{Garate2013}, whereas electron-magnon interactions favor it in the inverted regime ($m<0$). Magnon-magnon interactions, which we simplify here by a mean-field correction to the magnetization, can also induce complex effects such as topological phase transitions \cite{mook2021,lu2021}. Furthermore, whereas the electron-magnon and magnon-magnon interactions are controlled by the Curie temperature (ranging from 176 to 690 K in recently synthesized Weyl semimetals), the electron-phonon interactions are rather governed by the Debye temperature (typically, 300-400 K in most metals). Consequently, we expect electron-magnon interactions to dominate over electron-phonon interactions in low Curie temperature Weyl semimetals, such as $\mathrm{Co_3Sn_2S_2}$ ($T_c$=176 K), whereas in (Ni, Pt)MnSb ($T_c$=560-660 K) and $\mathrm{Co_2Mn(Ga,Al)}$ ($T_c$=690 K), both effects should affect the topology. Finally, the interaction between phonons and magnons can substantially affect the topology of magnons \cite{Zhang2019}, as recently reported in VI$_3$ monolayer \cite{Zhang2021} and Cu$_3$TeO$_6$ \cite{Chen2022}. Although our model only considers magnons with a trivial topology, the magnon spectrum is likely renormalized by their interaction with phonons. This effect is beyond the scope of the present work, but certainly constitutes an interesting research direction for the proper description of the anomalous and thermal Hall effects of Weyl semimetals. We point out that Weyl points can exist in non-collinear antiferromagnets $\mathrm{MnX_3}$ \cite{kuroda2017evidence,nayak2016} and altermagnetic CrSb \cite{lu2024observation,li2024topological}, suggesting that chiral magnons might get involved. 

Beyond the anomalous Hall effect, we expect that any physical properties associated with interband transitions could serve as a probe for the electron-magnon interaction, such as, but not limited to, the Nernst effect and the orbital magnetization. For instance, the orbital magnetization \cite{Thonhauser2005,Shi2007} is inversely proportional to the energy difference between bands $\sim1/(\varepsilon_n-\varepsilon_m)$. Thus, we expect it not only to increase near the Weyl points, similar to the anomalous Hall effect, but also to depend strongly on the relative positioning of the bands. This change of sign could be a signature of temperature-induced band inversion, absent in the anomalous Hall effect signal.

\textit{Data availability}
The data that support the findings of this article are available on Zenodo.

\begin{acknowledgments}
K. S. thanks Diego García Ovalle and Armando Pezo for fruitful discussions. K.S. and A.M. acknowledge support from the Excellence Initiative of Aix-Marseille Université–A*Midex, a French Investissements d’Avenir” program, as well as the grant ANR-23-CE09-0034-03 “NEXT” of the French Agence Nationale de la Recherche.
\end{acknowledgments}

\bibliography{bibliography}

\end{document}